\documentclass[sn-mathphys,Numbered]{sn-jnl}

\usepackage{graphicx}
\usepackage{multirow}
\usepackage{amsmath,amssymb,amsfonts}
\usepackage{amsthm}
\usepackage{mathrsfs}
\usepackage[title]{appendix}
\usepackage{xcolor}
\usepackage{textcomp}
\usepackage{manyfoot}
\usepackage{booktabs}
\usepackage{algorithm}
\usepackage{algorithmicx}
\usepackage[noend]{algpseudocode}
\usepackage{listings}
\usepackage{braket}
\usepackage{array}
\usepackage{caption}
\usepackage{subcaption}

\newcommand{\pluseq}{\mathrel{+}=}
\newcommand\figsize{0.85}
\newcommand\subfigsize{0.45}
\graphicspath{ {figures/} }

\usepackage{color}

\begin{document}

\title[Introducing UNIQuE: The Unconventional Noiseless Intermediate Quantum Emulator]{Introducing UNIQuE: The Unconventional Noiseless Intermediate Quantum Emulator}

\author*[1,2]{\fnm{Reece} \sur{Robertson}}\email{reecerobertson@umbc.edu}

\author[3]{\fnm{Dan} \sur{Ventura}}\email{ventura@cs.byu.edu}
\equalcont{These authors contributed equally to this work.}

\affil[1]{\orgdiv{Computer Science Department}, \orgname{University of Maryland, Baltimore County}, \orgaddress{\street{1000 Hilltop Circle}, \city{Baltimore}, \postcode{21250}, \state{Maryland}, \country{United States}}}

\affil[2]{\orgdiv{Physics Department}, \orgname{University of Maryland, Baltimore County}, \orgaddress{\street{1000 Hilltop Circle}, \city{Baltimore}, \postcode{21250}, \state{Maryland}, \country{United States}}}

\affil[3]{\orgdiv{Computer Science}, \orgname{Brigham Young University}, \orgaddress{\city{Provo}, \postcode{84602}, \state{Utah}, \country{United states}}}

\abstract{
    We implement the first open-source quantum computing emulator that includes arithmetic operations, the quantum Fourier transform, and quantum phase estimation.
    The emulator provides significant savings in both temporal and spatial resources compared to simulation, and these computational advantages are verified through comparison to the Intel Quantum Simulator.
    We also demonstrate how to use the emulator to implement Shor's algorithm and use it to solve a nontrivial factoring problem.
    This demonstrates that emulation can make quantum computing more accessible than simulation or noisy hardware by allowing researchers to study the behavior of algorithms on large problems in a noiseless environment.
}

\keywords{quantum computing, emulation, simulation}

\maketitle
\section{Introduction}

It is well-known that quantum computation harnesses entanglement and superposition to perform probabilistic calculation that results in super-classical advantage for some problems. 
However, current engineering constraints limit the capacities of today's quantum computers.
They are plagued by problems of limited size, limited connectivity, and nontrivial probability for errors \cite{errors}.
When these limitations are overcome, quantum algorithms will solve problems that are currently intractable.
Unfortunately, however, today's quantum computers have not yet achieved quantum utility.
To facilitate the development of quantum algorithms while the engineering issues are sorted out, researchers employ quantum computational \emph{simulation}, which allows classical computers to behave like quantum computers on a small scale.
These simulators are beneficial but, as demonstrated in this paper, there is an alternative approach called quantum computational \emph{emulation} that provides significant advantage over simulation.

The state of $n$ qubits can be represented mathematically by a state vector $\ket{\psi}$ of $2^n$ complex amplitudes.
Let $\ket{\psi_0}$ be the initial state of the qubits.
An algorithm on these qubits is represented by a sequence of $T$ operations that transforms $\ket{\psi_0}$ into $\ket{\psi_T}$.
Each of these $T$ operations can be represented by a $2^n\times2^n$ complex unitary matrix $M_t$, for all $t\in[1,T]$ \cite{khalid}.
A (discrete time) state vector simulator evolves the state of the qubits by multiplying $\ket{\psi_t}$ by $M_t$ for each $t$: $\ket{\psi_{t+1}} = M_{t}\ket{\psi_{t}}$.
This simulates exactly what would occur on a noiseless quantum computer and returns the final state vector $\ket{\psi_T}$ representing the state of the qubits at the conclusion of the algorithm.

Quantum computing emulation is a fundamentally different concept.
Rather than faithfully modeling each step $t$ of a quantum algorithm, an emulator \emph{only} returns the final state vector $\ket{\psi_T}$.
As such, a quantum emulator can abstract an entire algorithm into an optimized classical function.
For certain algorithms, emulation obtains temporal and/or spatial complexity lower than that for simulation.
Häner et al. identified three categories of functions for which an emulator attains these advantages: arithmetic operations (i.e., addition, multiplication, and exponentiation), the quantum Fourier transform, and quantum phase estimation \cite{emulator}.

We note that other's have used the term \textit{emulation} to refer to using specialized hardware for quantum computations.
Our work here is different; we are not using specialized hardware, but rather abstracting quantum algorithms into specialized classical functions \cite{khalid, fpga, fpga2, fpga3}.
Another paper uses \textit{emulator} to refer to a simulator coupled to a graphical user interface \cite{deraedt}, while still another uses the term to denote a quantum algorithm that implements a unitary operator \cite{lloyd}.

One additional point to note is that after measurement, a single shot of a quantum algorithm returns a binary string.
The results from repeated iteration of the algorithm can be interpreted as approximating a distribution over the space of possible output bitstrings (correlated with the wave function coefficients).
Hence, we consider the output of a quantum algorithm to be a distribution over a space of bitstrings.
This means that not all classical analogues to a quantum algorithm can be considered an emulation of that algorithm; rather, for a classical algorithm to emulate a quantum algorithm, it must return the same distribution over the space of binary strings that (repeated iteration of) the noiseless quantum algorithm would return.
Consequently, all functions employed by our emulator in the discussion that follows are both optimized for performance and return the same probability distributions that the ideal quantum version of the algorithm returns.

This paper introduces the Unconventional Noiseless Intermediate Quantum Emulator (UNIQuE)---the first open-source quantum emulator after the manner proposed by Häner et al.
UNIQuE can add, multiply, and exponentiate two superpositions of integers, as well as perform the quantum Fourier transform (and its inverse) and quantum phase estimation.
UNIQuE was tested against an existing state-of-the-art simulator to verify computational speedup.
During these comparisons UNIQuE also displayed appreciable spatial savings, which were improved with the use of sparse matrix operations where possible.
In addition, the functions of UNIQuE were used to perform Shor's factoring algorithm, which demonstrates the practical applicability of this software.
UNIQuE can be found on GitHub at \url{https://github.com/reecejrobertson/UNIQuE}.

\section{Mathematical Background} \label{math}

Recall that an $n$ quantum state vector is given by
\begin{equation}
  \sum_{i=0}^{2^n-1}\alpha_i\ket{i} = \left( {\begin{array}{c}
    \alpha_0 \\
    \alpha_1 \\
    \vdots   \\
    \alpha_{2^n-1}
  \end{array} } \right),
  \label{superposition}
\end{equation}
with the constraint that $\alpha_0,\cdots,\alpha_{2^n-1}\in\mathbb{C}$ and satisfy
\begin{equation}
    \sum_{i=0}^{2^n-1} |a_i|^2=1.
    \label{constraint}
\end{equation}
Any arbitrary vector can be normalized to satisfy (\ref{constraint}) by dividing by the root of the sum of the squares, that is,
\begin{equation}
  \mathcal{N}\left(\sum_{i=0}^{2^n-1}\alpha_i\ket{i}\right) = \frac{\sum_{i=0}^{2^n-1}\alpha_i\ket{i}}{\sqrt{\sum_{i=0}^{2^n-1} |a_i|^2}},
  \label{normalize}
\end{equation}
where $\mathcal{N}$ is the normalization function.

\subsection{Arithmetic Operations} \label{arithmetic_math}

The arithmetic operations of addition, multiplication, and exponentiation can be defined to operate on quantum states.
For each operation, a quantum computer receives two state vectors representing (potentially) several integers as input, and returns their sum, product, or the first raised to the power of the second, as the case may be.
For example, here is a concrete example of addition for a two qubit system:
\begin{equation*}
    \ket{\alpha} + \ket{\beta} = 
    \begin{pmatrix}
        \alpha_0 \\
        \alpha_1
    \end{pmatrix}
    +
    \begin{pmatrix}
        \beta_0 \\
        \beta_1
    \end{pmatrix}
    =
    \begin{pmatrix}
        \alpha_0\beta_0 \\
        \alpha_0\beta_1 + \alpha_1\beta_0 \\
        \alpha_1\beta_1 \\
        0
    \end{pmatrix}
    = \ket{\gamma}.
\end{equation*}

\begin{figure}[t]
\begin{algorithm}[H]
    \caption{Addition}\label{add}
    \begin{algorithmic}[1]
        \Require $\text{State vector } \ket{\alpha},\text{ state vector } \ket{\beta}$
        \Procedure{Add}{$\ket{\alpha}, \ket{\beta}$}:
            \State $N_{\alpha} \gets \text{length}(\ket{\alpha})$
            \State $N_{\beta} \gets \text{length}(\ket{\beta})$
            \State $\ket{\gamma} \gets \ket{0}\text{of length } 2\times\max(N_{\alpha}, N_{\beta})$
            \For{$i \gets 0, 1, ..., N_{\alpha}-1$}
                \For{$j \gets 0, 1, ..., N_{\beta}-1$}
                    \State $\ket{\gamma}_{i + j} \pluseq \ket{\alpha}_i \times \ket{\beta}_j$
                \EndFor
            \EndFor
            \State $\ket{\gamma} \gets \text{normalize}(\ket{\gamma})$
            \State \Return{$\ket{\gamma}$}
        \EndProcedure
    \end{algorithmic}
\end{algorithm}
\end{figure}

Pseudocode for general addition is given in Algorithm~\ref{add}.
Given two input vectors $\ket{\alpha}$ and $\ket{\beta}$, the output vector $\ket{\gamma}$ is constructed as follows:
each index in $\ket{\alpha}$ is added to each index in $\ket{\beta}$; this sum identifies an index in $\ket{\gamma}$ (left hand side of line 7 with the index denoted by the subscript).
The value of this entry of $\ket{\gamma}$ is increased by the product of the corresponding values of $\ket{\alpha}$ and $\ket{\beta}$ (right hand side of line 7).
The result is normalized by (\ref{normalize}) to satisfy constraint (\ref{constraint}) (line 8).

An actual quantum computer adds $\ket{\alpha}$ and $\ket{\beta}$ using a sequence of operations that grows linearly with $m=n_{\alpha}+n_{\beta}$, where $n_{\alpha}$ and $n_{\beta}$ denote the number of qubits used to represent $\ket{\alpha}$ and $\ket{\beta}$ \cite{ripple_carry}.
In like manner, a state vector simulator multiplies the $2^m\times1$ output state vector by a $2^m\times2^m$ matrix for each operation that the quantum computer performs.
This yields a complexity of at least $O(2^{2m}m)$ for simulation.
To emulate addition, on the other hand, it is only necessary to perform the mapping given by Algorithm $\ref{add}$ directly which requires $O(2^m)$ operations.
Therefore, we expect to see an increasing advantage for emulation over simulation as $m$ increases.

Multiplication is very similar to addition.  For a simple 2-qubit example, we have
\begin{equation*}
    \ket{\alpha} \times \ket{\beta} = 
    \begin{pmatrix}
        \alpha_0 \\
        \alpha_1
    \end{pmatrix}
    \times
    \begin{pmatrix}
        \beta_0 \\
        \beta_1
    \end{pmatrix}
    =
    \begin{pmatrix}
        \alpha_0\beta_0 + \alpha_0\beta_1 + \alpha_1\beta_0 \\
        \alpha_1\beta_1
    \end{pmatrix}
    = \ket{\gamma}.
\end{equation*}
The general algorithm is identical to Algorithm~\ref{add} except that $\ket{\gamma}$ is of length $N_{\alpha} \times N_{\beta}$ on line 4, and $\ket{\gamma}_{i\times j}$ is accessed on line 7.
The complete multiplication algorithm is given in Algorithm~\ref{multiply} (Appendix~\ref{arithmetic_algs_appendix}). 
Likewise, a 2-qubit example of exponentiation is
\begin{equation*}
    \ket{\alpha}^{\ket{\beta}} =
    \begin{pmatrix}
        \alpha_0 \\
        \alpha_1
    \end{pmatrix}
    ^
    {\begin{pmatrix}
        \beta_0 \\
        \beta_1
    \end{pmatrix}}
    =
    \begin{pmatrix}
        \alpha_0\beta_1 \\
        \alpha_0\beta_0 + \alpha_1\beta_0 + \alpha_1\beta_1
    \end{pmatrix}
    = \ket{\gamma}.
\end{equation*}
Again, the general algorithm is identical to Algorithm~\ref{add}, except this time the length of $\ket{\gamma}$ is $N_{\alpha}^{N_{\beta}}$ and we access $\ket{\gamma}_{i^j}$ on line 7 (see Algorithm~\ref{exponentiate} in Appendix~\ref{arithmetic_algs_appendix}).

When it comes to multiplication and exponentiation, theoretically the emulation method still only requires $O(2^m)$ operations (in practice, this advantage depends upon how $\ket{\gamma}$ is initialized; see Section~\ref{performance}).
The simulation method, however, may require many more---multiplication is repeated addition and exponentiation is repeated multiplication.
Hence, standard multiplication requires $O(2^{2m}m^2)$ operations and standard exponentiation requires $O(2^{2m}m^3)$ operations.
In sum, we expect an advantage for emulation over simulation for all arithmetic operations as $m$ increases.

\subsection{Quantum Fourier Transform} \label{qft_math}

The quantum Fourier transform (QFT) is a unitary operator that takes an $n$ qubit input state $\ket{\alpha}$ of the form of (\ref{superposition}) and maps it to a state 
\begin{equation}
    \sum_{j=0}^{2^n-1}\beta_j\ket{j},
\end{equation}
where each of the $\beta_j$ are given by the classical discrete Fourier transform of the amplitudes of $\ket{\alpha}$ \cite{qft}:
\begin{equation}
    \beta_j=\frac{1}{\sqrt{2^n}}\sum_{j=0}^{2^n-1} \alpha_j e^{2\pi i j k/2^n}.
    \label{fft}
\end{equation}
One subtlety to note here is that (\ref{fft}) is actually the \emph{inverse} discrete Fourier transform.  UNIQuE also implements the inverse quantum Fourier transform, which is defined similarly to the QFT but where each $\beta_j$ is given by the standard discrete Fourier transform:
\begin{equation}
    \beta_j=\frac{1}{\sqrt{2^n}}\sum_{j=0}^{2^n-1} \alpha_j e^{-2\pi i j k/2^n}.
    \label{ifft}
\end{equation}

Because UNIQuE makes use of classical libraries, this function can be implemented in a single line, as shown in Algorithm~\ref{qft_code}.
Here ``IFFT'' represents the inverse fast Fourier transform, a celebrated classical function that implements (\ref{fft}) quickly \cite{fft}.\footnote{UNIQuE utilizes the \texttt{scipy.fftpack} package for both FFT and IFFT: \url{https://docs.scipy.org/doc/scipy/reference/fftpack.html}.}
The inverse quantum Fourier transform implementation is identical, except for the substitution of ``FFT,'' the standard fast Fourier transform, which implements (\ref{ifft}).

\begin{figure}[t]
\begin{algorithm}[H]
    \caption{Quantum Fourier Transform}\label{qft_code}
    \begin{algorithmic}[1]
        \Require $\text{State vector } \ket{\alpha}$
        \Procedure{QFT}{$\ket{\alpha}$}:
            \State \Return{$\text{normalize(IFFT}(\ket{\alpha}))$}
        \EndProcedure
    \end{algorithmic}
\end{algorithm}
\end{figure}

As outlined by Häner et al., the quantum Fourier transform algorithm on $n$ qubits requires $n$ single-qubit operations, and $n(n-1)/2$ two-qubit operations.
This means that to simulate the quantum Fourier transform requires $O(2^{n}n^2)$ total operations \cite{emulator}.
However, to perform (\ref{fft}) directly via the FFT requires only $O(2^{n}n)$ operations \cite{fft}
(the numbers are identical for the inverse quantum Fourier transform).
Therefore, once again we expect to see that emulation gains temporal advantage over simulation.

\subsection{Quantum Phase Estimation} \label{qpe_math}

\begin{figure}[t]
\begin{algorithm}[H]
    \caption{Quantum Phase Estimation}\label{qpe_code}
    \begin{algorithmic}[1]
        \Require $\text{NxN unitary matrix } U,\text{ eigenvector } \boldsymbol\phi, \text{ integer b}$
        \Procedure{QPE}{$U, \boldsymbol\phi, b$}:
            \State $\text{evecs} \gets \text{eigenvectors(U)}$
            \State $\text{evals} \gets \text{eigenvalues(U)}$
            \State $\text{z} \gets \text{evals}[\text{evecs}[\boldsymbol\phi]]$
            \State $\theta \gets \ln(z)/(2\pi i)$
            \State $r \gets \theta\times2^b$
            \State $r \gets \text{int}(\text{real}(r))$
            \State $\ket{\gamma} \gets \ket{0}\text{of length }2^b$
            \State $\ket{\gamma}_r \gets 1$
            \State \Return $\ket{\gamma}$
        \EndProcedure
    \end{algorithmic}
\end{algorithm}
\end{figure}

Quantum phase estimation (QPE) takes as input a $2^n\times2^n$ unitary matrix $U$, an eigenvector $\ket{\phi}$ encoded into a state vector of $n$ qubits, and a second state vector of $b$ qubits.
Because $U$ is unitary, all of its eigenvalues lie on the unit circle in the complex plane, and therefore the eigenvalue of $U$ associated with $\ket{\phi}$ is of the form
\begin{equation}
    z = e^{2\pi i\theta}.
    \label{phase}
\end{equation}
QPE estimates $\theta$ by selecting the one of the $2^b$ roots of unity closest to it \cite{qpe}.
Increasing $b$ yields a more precise estimate if it is not exact.

Algorithm~\ref{qpe_code} demonstrates how to emulate quantum phase estimation.
The algorithm works as follows: find the eigenvalue $z$ of $U$ associated with the eigenvector $\ket{\phi}$ (lines 2-4).
Use that eigenvalue to compute the phase $\theta$ from (\ref{phase}) (line 5).
Next, find the element $r$ of the $2^b$ roots of unity which is closest to $\theta$ by multiplying $\theta$ by $2^b$ (line 6).
Encode this into a state vector $s$ of size $2^b$ by using the real part of $r$ as the index of state vector $s$, whose amplitude is set to 1 (lines 7-9).

On a quantum computer, QPE consists of $b$ controlled unitary rotations followed by the IQFT.
Simulating the controlled unitary operations requires $O(2^{b}b)$ operations \cite{emulator}; combining this with the complexity of the IQFT gives a total complexity of ${O(2^{b+n}bn^2)}$.
An emulator, however, finds the eigenvalue of interest directly and encodes the associated $b$ bit phase estimation into an appropriately sized state vector.
This has a complexity of $O(2^{2.38n})$ using an optimized classical eigenvalue algorithm \cite{eigenvalues}, suggesting that emulation will outperform simulation at least when $n+b \geq 2.38n$.
Interestingly, the complexity of emulation theoretically does not depend upon $b$ (in practice dependence may arise for some representations of $\ket{\gamma}$), and hence we expect emulation to obtain exponential advantage over simulation as $b$ increases.

\section{Performance Results} \label{performance}

UNIQuE implements the three arithmetic operations, the quantum Fourier transform (and its inverse), quantum phase estimation, the normalization protocol (\ref{normalize}), and the quantum measurement protocol.
For normalization and the arithmetic operations UNIQuE contains both a non-sparse and sparse implementation.
The documentation for each function provided by UNIQuE can be found in Appendix~\ref{documentation}.

After UNIQuE was constructed it was tested against the Intel Quantum Simulator (Intel-QS; formerly called qHiPSTER) \cite{qhipster} to verify its advantage.
This simulator is open-source, and available at \url{https://github.com/iqusoft/intel-qs}.
All comparisons were performed on a supercomputer CPU with access to 2TB of RAM.

\subsection{Arithmetic Operations}\label{arithmetic_performance}

\begin{figure}[t]
\centering
\includegraphics[width=\figsize\linewidth]{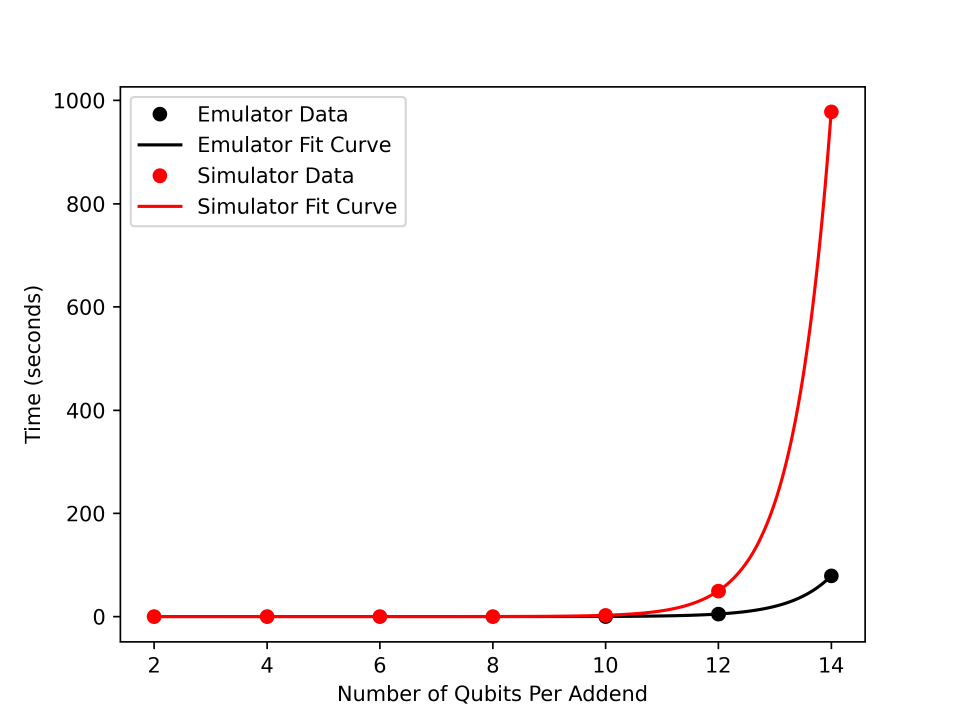}
\caption{
    A comparison of the addition algorithm between UNIQuE and Intel-QS.
    Both implementations scale exponentially, but Intel-QS scales with a larger exponent, demonstrating UNIQuE's temporal advantage for this operation.
}
\label{addition_figure}
\end{figure}

To perform the comparison between UNIQuE and Intel-QS on the addition operation, two state vectors of $n$ qubits each were initialized to random classical states, for $n\in[2,4,6,8,10,12,14]$.
The two numbers were added together, using the \texttt{add} method on UNIQuE and a ripple-carry algorithm \cite{ripple_carry} on Intel-QS.
The two results were compared to verify that they were identical in every case.
Additionally, each method was timed, and the average times over ten repetitions of this experiment are presented in Fig.~\ref{addition_figure}.

In Section~\ref{arithmetic_math} it is estimated that simulated addition scales as $O(2^{2n}n)$, while emulated addition scales as $O(2^n)$.
Experimentally, however, we see somewhat contrary results.
The Intel-QS result is best fit by the curve
\begin{equation*}
    y \approx (8.506 \times 10^{-7}) \times 2^{2.152n},
\end{equation*}
with a mean squared error of 0.012, while the best fit curve of the expected form is 
\begin{equation*}
    y \approx (1.789 \times 10^{-7}) \times n \times 2^{2.041n},
\end{equation*}
with an error of 0.032.
The UNIQuE result, on the other hand, is best fit by the curve
\begin{equation*}
    y \approx (6.802\times10^{-08}) \times n \times2^{1.880n},
\end{equation*}
with a mean squared error of $5.299 \times 10^{-6}$, while the best fit curve of the expected form is
\begin{equation}
    y \approx (3.232 \times 10^{-7}) \times 2^{1.992n},
    \label{em_addition}
\end{equation}
with an error of $4.584 \times 10^{-5}$.
While UNIQuE does scale with a smaller exponent, the curve of best fit for UNIQuE includes an additional multiplicative factor of $n$, while the curve for Intel-QS does not.
However, the difference in error between the two forms is very small for both UNIQuE and simulator curves, indicating that minor variability in the run time may significantly impact the curve of best fit.
Therefore, comparing larger systems may be necessary to confirm the expected scaling patterns.\footnote{Unfortunately, even utilizing 2TB of RAM it is not possible to simulate more than 14 qubits per addend on Intel-QS.}

\begin{figure}[t]
    \centering
    \begin{subfigure}{\subfigsize\textwidth}
        \centering
        \includegraphics[width=\figsize\linewidth]{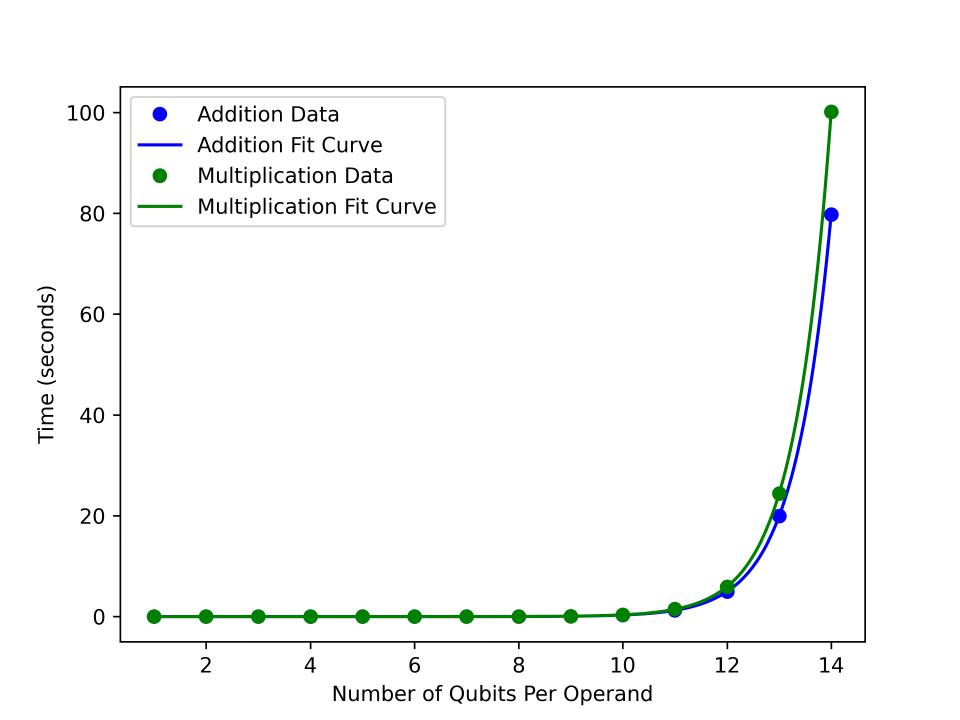}
        \caption{
            A comparison of non-sparse emulated addition and multiplication.
        }
        \label{multiplication_figure}
    \end{subfigure}
    \begin{subfigure}{\subfigsize\textwidth}
        \centering
        \includegraphics[width=\figsize\linewidth]{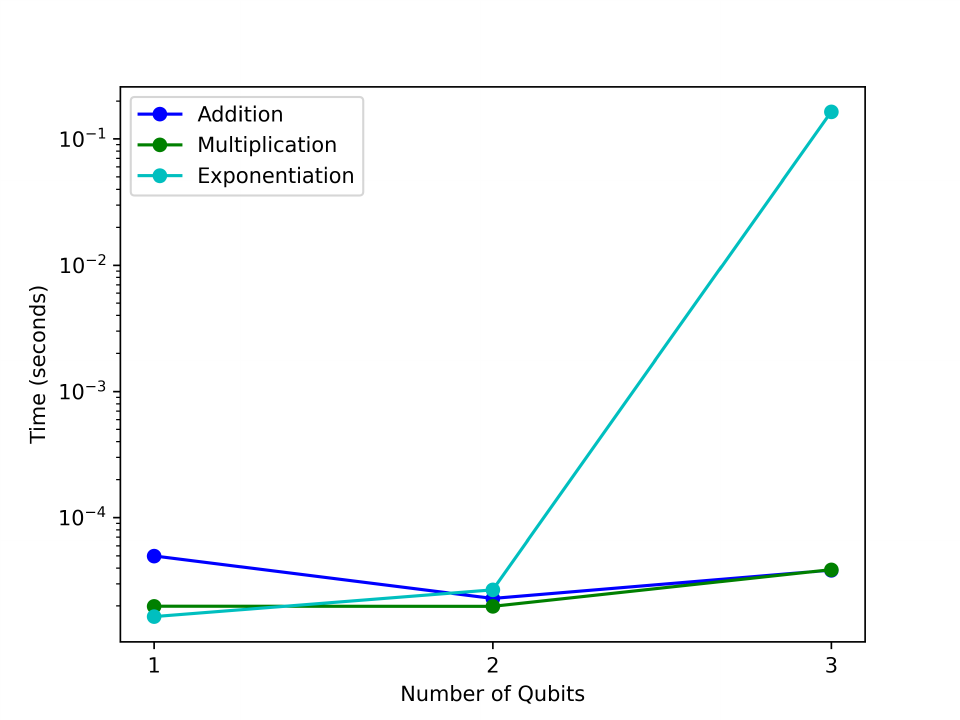}
        \caption{
            Emulated addition, multiplication, and exponentiation.
            Note the log scale.
        }
        \label{exponentiation_figure}
    \end{subfigure}
    \caption{
        The non-sparse UNIQuE implementation of addition, multiplication, and exponentiation.
        Observe that addition and multiplication obtain a comparable exponential scaling, however exponentiation scales super-exponentially.
        Moreover, memory constraints on the supercomputer prohibit non-sparse exponentiation on state vectors representing more than three qubits.
    }
\end{figure}

Later in Section~\ref{arithmetic_math}, it is postulated that simulated multiplication and exponentiation should be no more expensive than addition.
This conjecture assumes that state vector size does not affect computational complexity, which in general depends upon the data structure utilized.
We tested this in the non-sparse setting by generating two random state vectors of size $2^n$ with $n\in[1,14]$, and then performing the non-sparse arithmetic operations on these vectors.
We repeated this ten times, with new random vectors for each iteration.
A comparison of the average speed of emulated multiplication against emulated addition is given in Fig.~\ref{multiplication_figure},
and a comparison of the speed of emulated exponentiation against the others is given in Fig.~\ref{exponentiation_figure}.\footnote{Note that these figures only compare the speed of the different emulated operations; simulated operations are not represented in either figure.}

As predicted, multiplication scales similarly to addition.
In fact, the emulated multiplication is best approximated by the curve
\begin{equation}
    y \approx (3.256 \times 10^{-7}) \times 2^{2.001n},
    \label{em_multiplication}
\end{equation}
which is nearly identical to the expected addition curve (Eq.~\ref{em_addition}).
However, when using a non-sparse method, exponentiation scales much worse than either addition or multiplication: the time and space requirements for emulating addition and multiplication scale exponentially in the number of qubits, while the time and space requirements for exponentiation scale super-exponentially.
As a result, performing exponentiation with any more that three qubits is beyond the capacity of our hardware.\footnote{With four qubits, there are $2^4=16$ values, which means the number of values required for emulating exponentiation is $16^{16} \gg 2 \times 10^{12} = 2\text{TB}$.}
This issue is remedied by the sparse version of this function, which is discussed next.

\begin{figure}[t]
    \centering
    \begin{subfigure}{\subfigsize\textwidth}
        \centering
        \includegraphics[width=\figsize\linewidth]{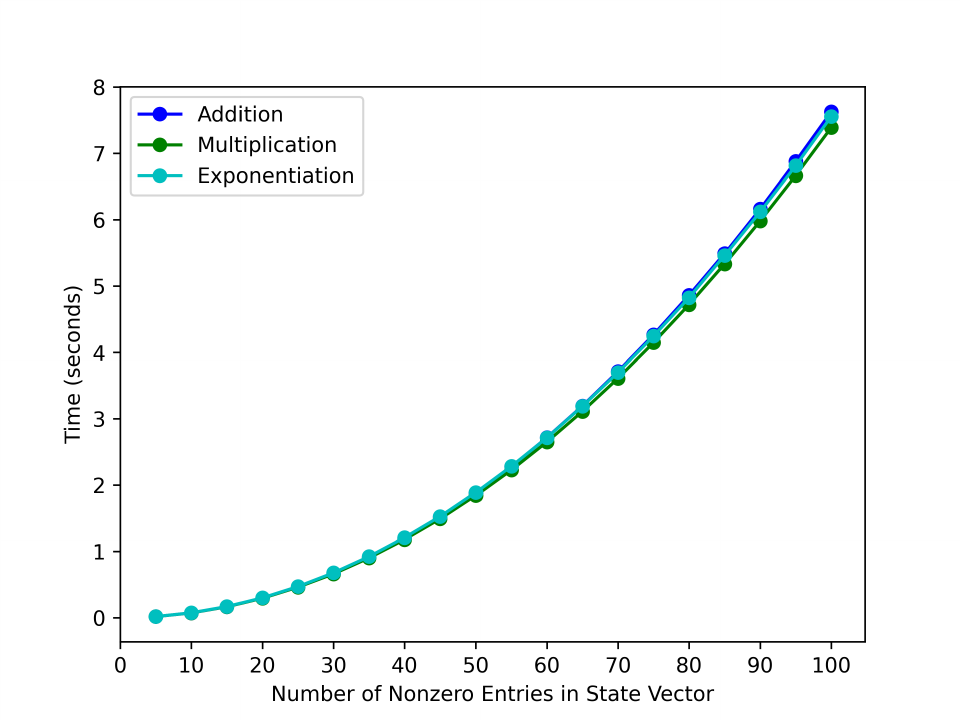}
        \caption{
            Ten qubit operands.
        }
        \label{sparse_states_figure}
    \end{subfigure}
    \begin{subfigure}{\subfigsize\textwidth}
        \centering
        \includegraphics[width=\figsize\linewidth]{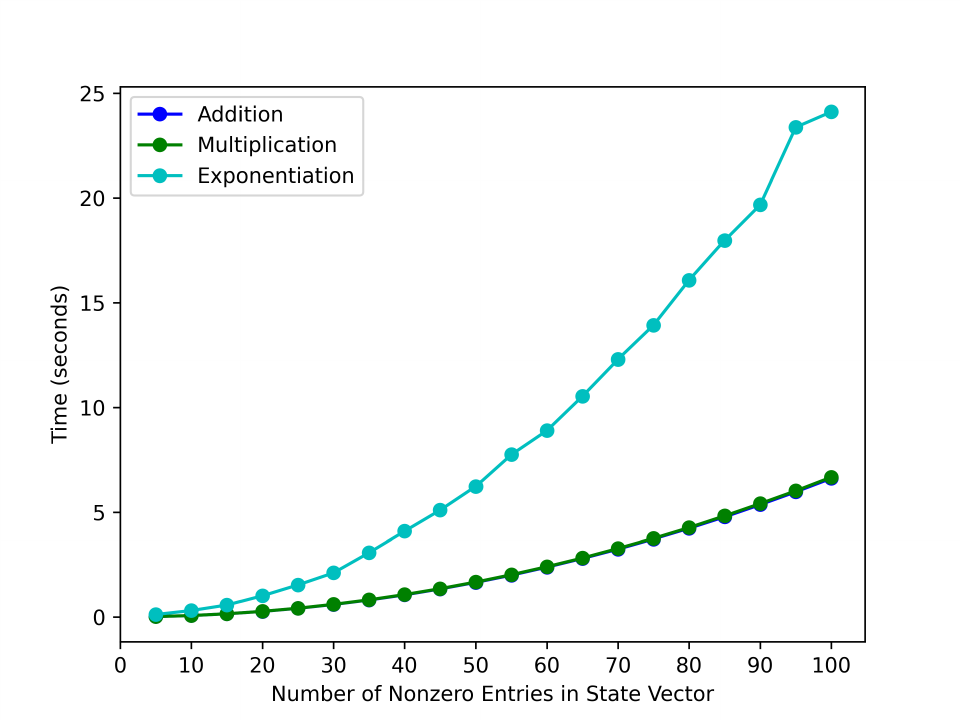}
        \caption{
            Twenty qubit operands.
        }
\label{sparse_states_second_figure}
    \end{subfigure}
    \caption{
        A comparison of sparse addition, multiplication, and exponentiation with the number of qubits in each operand held constant, while the number of nonzero entries in each state vector varies.
        For ten qubits all three operations scale nearly identically, as opposed to the non-sparse method where exponentiation scales drastically worse than the others.
        For twenty qubits the super-exponential nature of emulating exponentiation again becomes apparent.
    }
\end{figure}

When state vectors contain many empty entries, emulation using a sparse matrix package can be beneficial.
Fig.~\ref{sparse_states_figure} demonstrates emulation using sparse operations on two ten-qubit state vectors that contain a variable number of nonzero entries.
The locations and values of these nonzero entries were selected randomly for each of ten trials.
The average time required for each operation is plotted as a function of the number of nonzero entries in each operand.
For comparison, the non-sparse method performed ten-qubit addition and multiplication in 0.311 and 0.338 seconds respectively.
Hence, the sparse framework is advantageous for these operations at this scale if they contain ~60 or fewer nonzero entries (and it is always advantageous for exponentiation).

\begin{figure}[t]
    \centering
    \includegraphics[width=\figsize\linewidth]{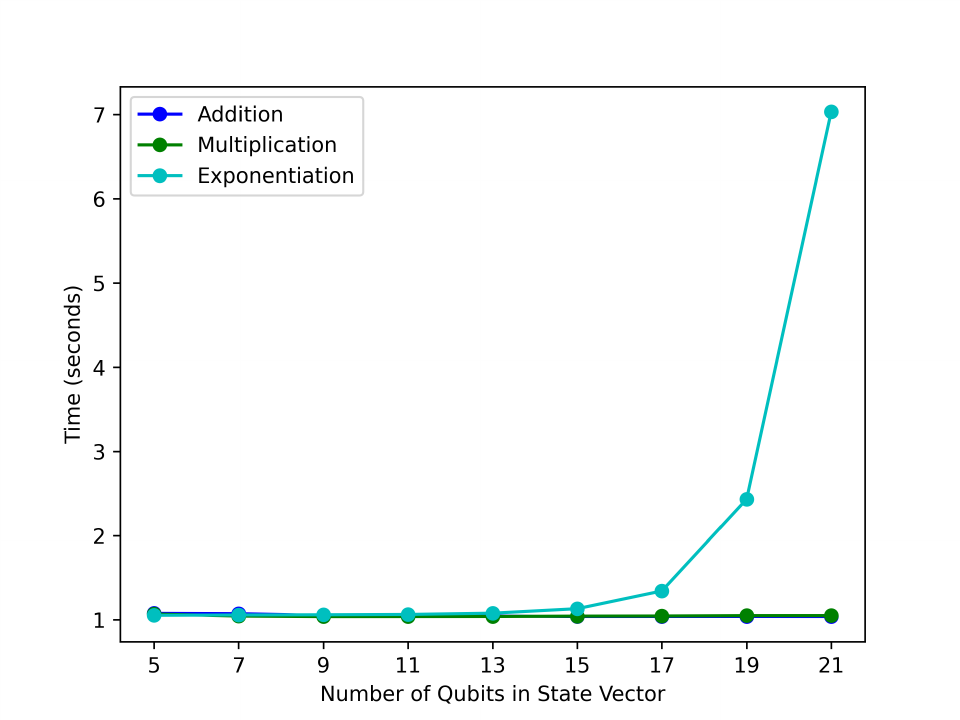}
    \caption{
        A comparison of sparse addition, multiplication, and exponentiation where the number of nonzero entries in each state vector is constant, while the number of qubits in each operand varies.
        Here again it can be seen that exponentiation scales worse than addition and multiplication.
    }
    \label{sparse_qubits_figure}
\end{figure}

The same experiment was conducted with sparse state vectors containing values for twenty qubits (Fig.~\ref{sparse_states_second_figure}).
The standard version of these operations would take approximately $3.181\times10^5$ seconds for addition and $3.630\times10^5$ seconds for multiplication, as computed by equations (\ref{em_addition}) and (\ref{em_multiplication}) respectively.
It is infeasible to attempt these operations using the standard method, and so here the sparse framework makes all the difference.

Fig.~\ref{sparse_qubits_figure} shows UNIQuE's performance operating on two sparse state vectors with a variable number of qubits that always contain thirty nonzero entries (again random values in random locations).
The average time for each operation over ten iterations of the experiment is shown.
As expected, exponentiation does not benefit as much from the sparse framework; however, here again sparse exponentiation can emulate larger circuits than is otherwise possible (compare to Fig.~\ref{exponentiation_figure}).

\begin{figure}[t]
    \centering
    \includegraphics[width=\figsize\linewidth]{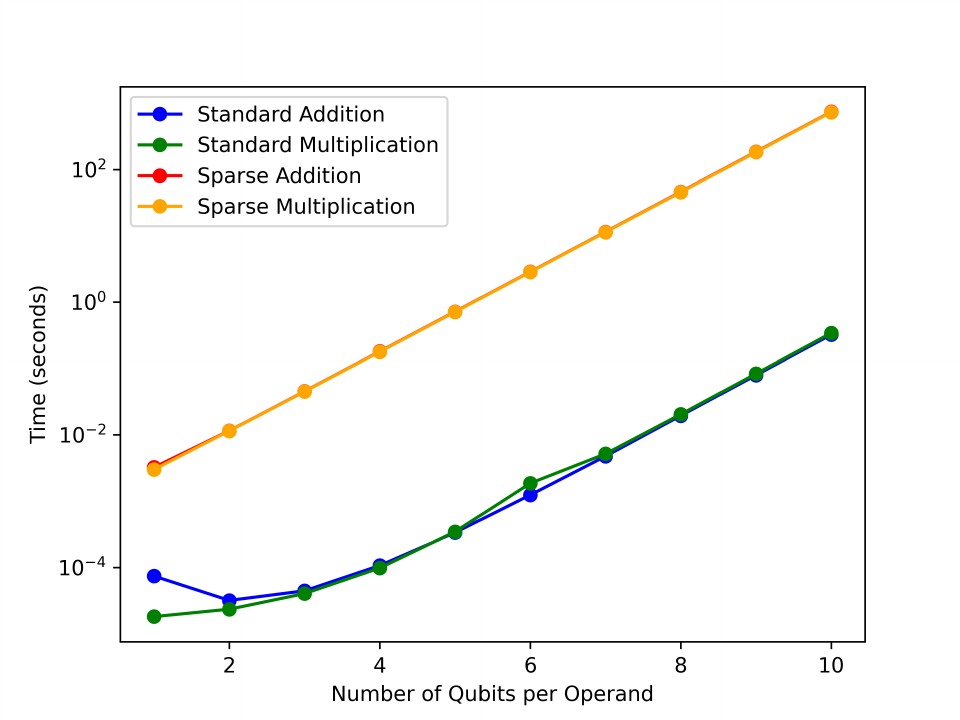}
    \caption{
        A comparison of standard and sparse emulated addition and multiplication on dense state vectors.
        Both operations scale identically; however, the sparse framework requires significant additional overhead.
        Note the log scale.
    }
    \label{dok_mult}
\end{figure}

Finally, Fig.~\ref{dok_mult} compares the standard and sparse implementations for emulating addition and multiplication on dense state vectors.
The run time for the sparse operations (the upper two curves in the figure) scale with time complexities of
\begin{equation*}
    y \approx (7.437\times10^{-4}) \times 2^{2.001n}
\end{equation*}
for addition and
\begin{equation*}
    y \approx (7.319\times10^{-4}) \times 2^{2.002n}
\end{equation*}
for multiplication.
Comparing these to to equations (\ref{em_addition}) and (\ref{em_multiplication}) it is clear that the sparse implementation incurs a significant overhead for emulating dense state vectors.

There are several conclusions to draw from these results with a sparse implementation of the arithmetic operations.
First, for sparse state vectors a sparse emulation yields dramatic spatial savings---we demonstrated an increase of 17 qubits over the traditional method for exponentiation, and we anticipate that a larger gain could be achieved.
Second, for state vectors with few nonzero entries the runtime of sparse emulation is comparable to the runtime of standard emulation; however, the runtime of sparse emulation degrades as the number of nontrivial entries increases.
Finally, while addition and multiplication maintain near constant runtime as the number of qubits increases, the runtime of exponentiation increases exponentially with qubit number.

To sum up, emulating quantum arithmetic operations yields temporal speedup over classical methods, along with appreciable spatial savings.
Using a sparse formulation of these problems can yield significant spatial savings and additional speedup if the number of nonzero entries is sufficiently small.
Combined, this means that UNIQuE can solve problems more quickly than simulators can, and it can also solve much larger sparse problems than simulators can handle.

\subsection{Quantum Fourier Transform}

\begin{figure}[t]
    \centering
    \includegraphics[width=\figsize\linewidth]{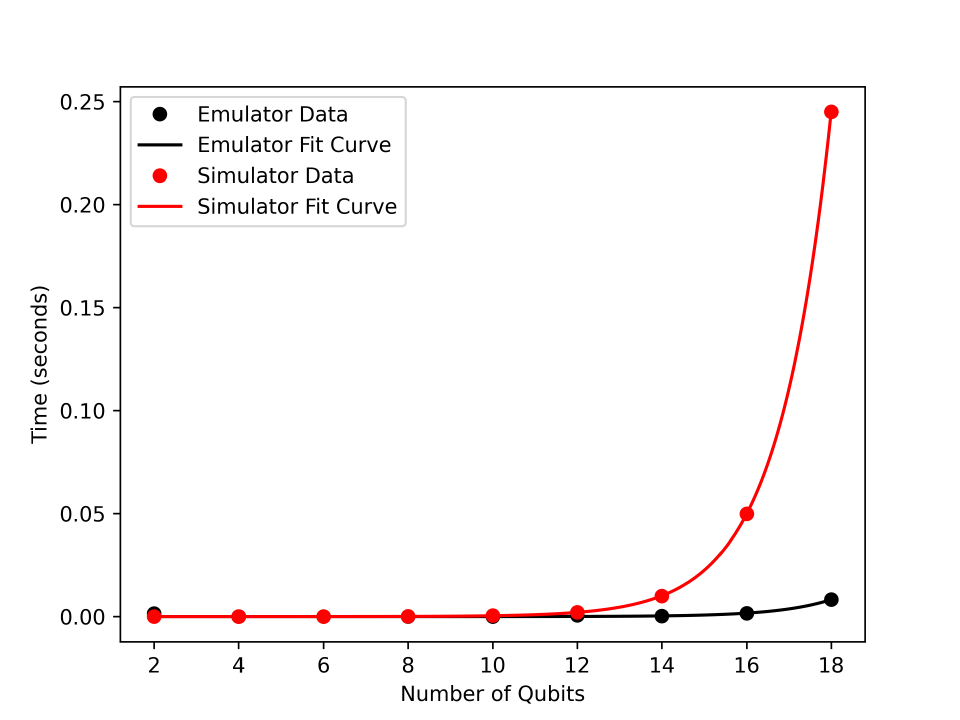}
    \caption{
        A comparison of the Quantum Fourier Transform between our UNIQuE and Intel-QS.
        The complexity of both implementations scales exponentially as qubit count increases; however UNIQuE's scaling coefficient is smaller.
    }
    \label{qft_figure}
\end{figure}

To determine the advantage of emulation over simulation for the quantum Fourier transform, several random state vectors representing $n$ qubits were generated for even $n$ between 2 and 18.
The QFT was then applied to these vectors using the \texttt{qft} function on UNIQuE and Nielsen and Chuang's presentation of the algorithm on Intel-QS \cite{quantum_bible}.
The output of UNIQuE and Intel-QS agreed in every case.
Once again, each operation was timed and the average time over ten iterations is given in Fig.~\ref{qft_figure}.

The results from UNIQuE are best approximated by the curve
\begin{equation*}
    y \approx (4.950 \times 10^{-9}) \times 2^{1.148n},
\end{equation*}
while the results from Intel-QS are best approximated by
\begin{equation*}
    y \approx (1.454 \times 10^{-7}) \times 2^{1.149n}.
\end{equation*}
Hence both operations scale with a comparable exponential increase in complexity; however UNIQuE has a significantly smaller leading coefficient than Intel-QS.
While our fit curves are not of the form predicted by Section~\ref{qft_math}, it is clear that UNIQuE maintains exponential advantage over Intel-QS.\footnote{The results were not well-approximated by a curve of the expected form which included a factor of $n^x$.}

Häner et al. found that, contrary to their complexity analysis, the speedup of QFT emulation diminished as $n$ increased for $n\in\{28, 30, 32, 34, 36\}$.
Our results do not demonstrate such behavior, but instead suggest that emulation maintains a constant speedup over simulation.
However, given our computational resources we could only simulate 18 qubits.\footnote{It should be noted that we emulated up to 30 qubits for this operation.}
This begs the question of how UNIQuE will perform relative to Intel-QS for larger values of $n$.
It could be that advantage will degrade as Häner et al. found, or that it will retain its constant speedup, or that the predicted increasing speedup will emerge.
In spite of this uncertainty, it is clear that emulation offers a performance advantage for the QFT at intermediate scales.

\subsection{Quantum Phase Estimation}

Quantum phase estimation demonstrates the most dramatic divergence between emulation and simulation.
To show this, we generated several $2\times2$ unitary matrices of the form
\begin{equation}
    U = \begin{pmatrix}
        1 & 0 \\
        0 & e^{iz}
    \end{pmatrix},
    \label{Umatrix}
\end{equation}
with $z$ drawn from the uniform distribution on $[0,1]$.
Regardless of the value of $z$, $U$ has an eigenvector
\begin{equation*}
    \ket{\phi}=\begin{pmatrix}
        0 \\
        1
    \end{pmatrix}.
\end{equation*}
Then $U$, $\ket{\phi}$, and a precision $b\in\mathbb{N}$ were passed to UNIQuE's \texttt{qpe} function to estimate the phase $\theta=z/2\pi$.
For comparison, $\ket{\phi}$ was encoded (via a \texttt{NOT} gate) into the last qubit of a $b+1$ qubit state vector on Intel-QS, and $\theta$ was estimated using Nielsen and Chuang's presentation of QPE \cite{quantum_bible}.
Both estimations were compared for consistency; UNIQuE and Intel-QS agreed in every case, up to global phase.\footnote{While UNIQuE does not preserve global phase, Intel-QS does. Because global phase cannot be detected in a real quantum measurement, the two outputs are functionally equivalent.}

\begin{figure}[t]
    \centering
    \includegraphics[width=\figsize\linewidth]{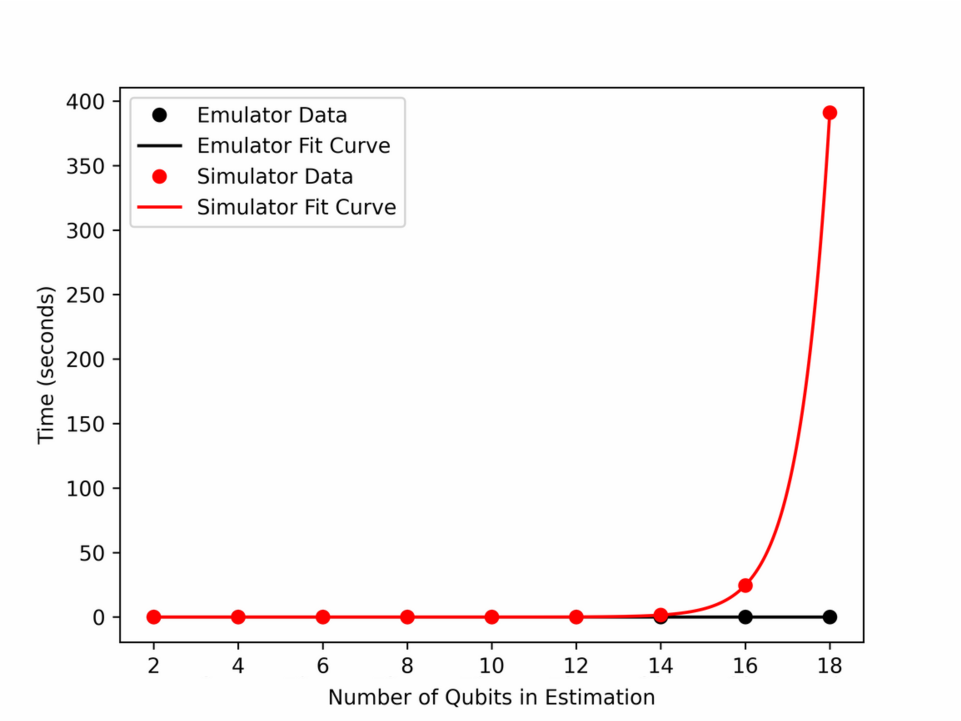}
    \caption{
        Quantum phase estimation performed with UNIQuE and Intel-QS.
        A $2\times2$ matrix is used for each estimation; here ``number of qubits'' on the $x$-axis refers to the estimation precision, $b$.
        QPE runtime on Intel-QS scales exponentially, while runtime on UNIQuE is constant.
    }
    \label{qpe_figure}
\end{figure}

Our first set of experiments evaluated the effect of estimation precision, $b$, on runtime.
We repeated the above process ten times for the even values of $b$ between 2 and 18, using a different random $U$ matrix for each repetition.
The average computational times for both Intel-QS and UNIQuE are plotted in Fig.~\ref{qpe_figure} as a function of $b$.
Here again Intel-QS shows an exponentially increasing complexity; its data is approximated by the curve
\begin{equation*}
    y \approx 6.067 \times 2^{1.995b}.
\end{equation*}
This is consistent with the calculations of Section~\ref{qpe_math}, which predicted that the time required to simulate QPE is exponential in both the number of qubits used for $U$ ($n$), and the number of qubits used for the estimation ($b$).

\begin{figure}[t]
    \centering
    \includegraphics[width=\figsize\linewidth]{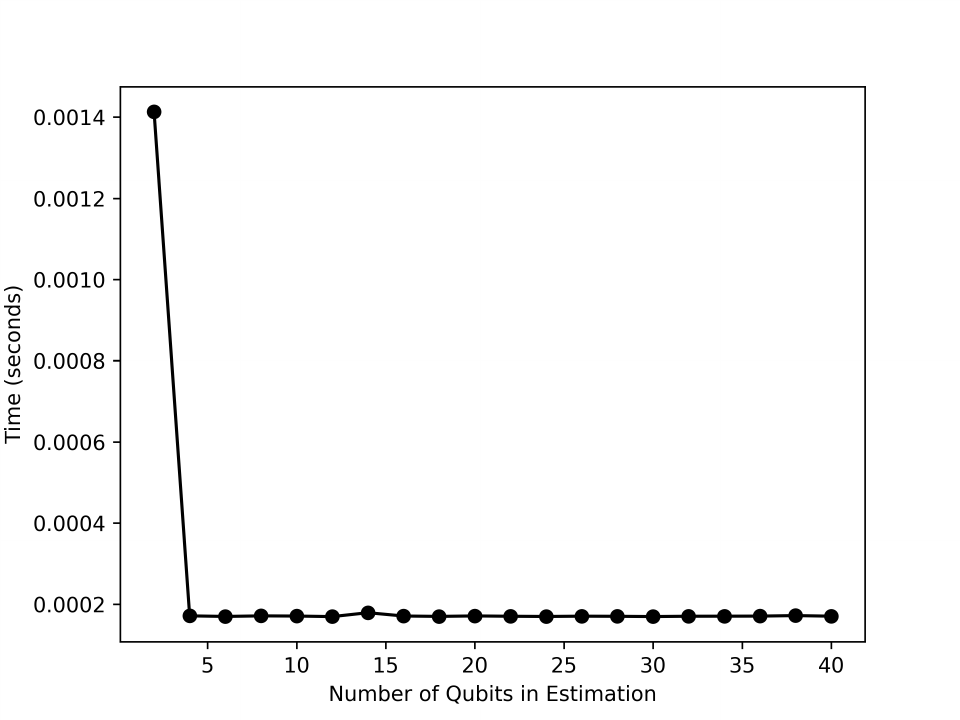}
    \caption{
        Quantum phase estimation performed on UNIQuE alone.
        The $x$-axis corresponds to the estimation precision, $b$, of the eigenvalue of a $2\times2$ matrix.
        The runtime of this operation is constant in $b$.
    }
    \label{em_qpe_figure}
\end{figure}

For UNIQuE, however, the story is much different.
It is predicted in Section~\ref{qpe_math} that the emulation complexity is only exponential in $n$.
Here $n$ is constant, so we expect runtime to be likewise constant.
This appears to be true in Fig.~\ref{qpe_figure}.
To investigate further, we performed QPE with UNIQuE on 100 random $2\times2$ unitary matrices of form (\ref{Umatrix}) with $b\in[1,40]$.
The average runtime is plotted in Fig.~\ref{em_qpe_figure} as a function of $b$.
This concretely demonstrates that the complexity of QPE on UNIQuE is independent of $b$.  Observe that with 40 qubits there are $2^{40}$ roots of unity with which to estimate the angle $\theta$ of (\ref{phase}).
This provides an absurd level of accuracy, far greater than required for most computations (especially considering the current scarcity of qubits as a resource).
Thus, UNIQuE can achieve extraordinary accuracy with no extra cost in terms of computational complexity.

\begin{figure}[t]
    \centering
    \includegraphics[width=\figsize\linewidth]{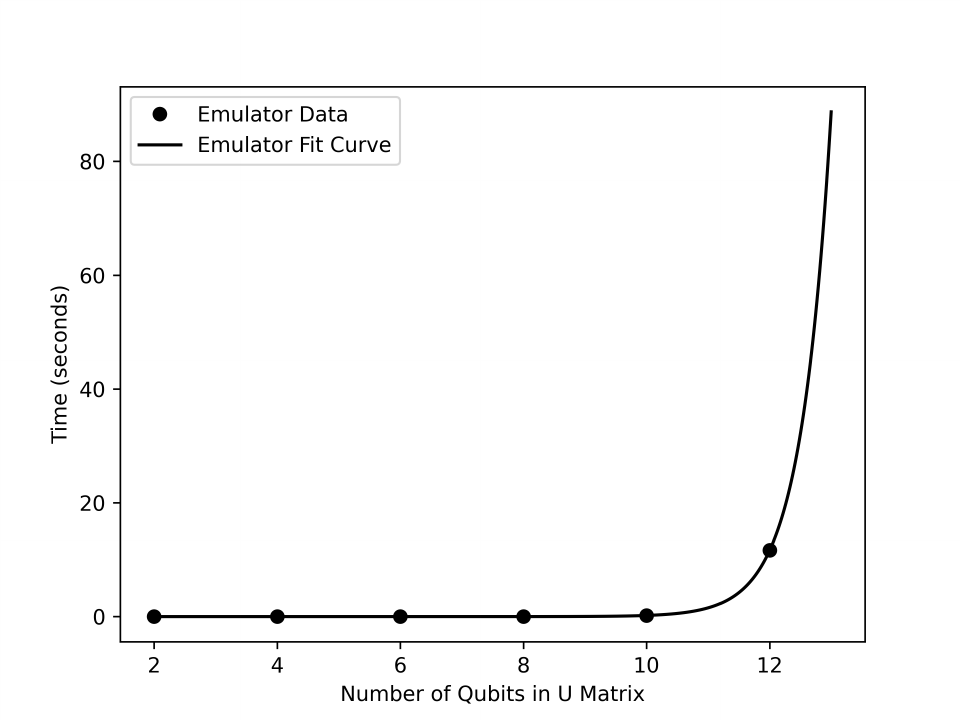}
    \caption{
        Quantum phase estimation performed on UNIQuE alone as a function of $n$ (the number of qubits in the $U$ matrix).
        As predicted, the complexity of emulating QPE scales exponentially in $n$.
    }
    \label{em_qpe2_figure}
\end{figure}

Our second set of experiments validates that the complexity of QPE on UNIQuE scales exponentially as $n$ increases.
The result is shown in Fig.~\ref{em_qpe2_figure}; the curve is best approximated by
\begin{equation*}
    y \approx (3.111 \times 10^{-10}) \times 2^{2.927n}.
\end{equation*}
This is as expected, and given the computation in Section~\ref{qpe_math} it is expected that simulated QPE will scale similarly in $n$.
However, it should be noted that Intel-QS (and many others) do not permit a controlled operation that targets more than a single qubit.
This means that to successfully simulate QPE using a $U$ matrix that spans multiple qubits, one must manually decompose the successive applications of powers of $U$ to smaller operations.
This can be a complicated process---UNIQuE obviates the need for such decompositions.
Thus, UNIQuE makes these QPE more accessible on at least two accounts: it allows users to run it quickly with high precision, and it removes the necessity of manual quantum gate decompositions.

\subsection{Spatial Advantage}\label{spatial_savings}

Recall from Section~\ref{arithmetic_performance} a sparse emulation of the arithmetic operations yields dramatic spatial savings for sufficiently sparse state vectors.
Explicitly we showed exponentiation of two 20 qubit state vectors, which requires $2^{20971520} \approx 10^{6300000}$ values when simulated.
It is impossible to store such a large state vector on any classical hardware; thus sparse emulation makes the problem tractable.

Unfortunately QFT does not lend itself well to a sparse implementation; in general sparse states are mapped to dense states by the QFT.
For this reason we did not implement a sparse QFT function.
For QPE, however, there is always exactly one nontrivial entry in the output state vector.
This is so naturally sparse that we chose to always returns a sparse data type for QPE.

However, even disregarding sparse implementations, UNIQuE offers an appreciable spatial advantage over Intel-QS.
On a laptop with 8GB of RAM, UNIQuE achieved a maximum of $n=29$ qubits across all operations, while Intel-QS achieved only $n=18$ qubits.
Here both UNIQuE and Intel-QS store the entire $2^n$ state vector, but UNIQuE saves space by avoiding all $2^n\times2^n$ matrix multiplications.

\section{Practical Application} \label{application}

To demonstrate the practical applicability of UNIQuE, we used it to run the celebrated factoring algorithm discovered by Peter Shor.
This algorithm takes as input two co-prime integers $X$ and $a$.
A successful execution returns the smallest number $r$ such that 
\[
a^r=1 \mod X.
\]
Given $r$, it follows that
\[
(a^r-1)=0 \mod X,
\]
hence $X$ divides $(a^r-1)$.  Moreover, if $r$ is even then
\begin{equation}
a^r-1=(a^{r/2}-1)(a^{r/2}+1),
\label{factors}
\end{equation}
and it is likely that the greatest common divisor of $X$ and $a^{r/2}\pm1$ is a proper factor of $X$ \cite{shor, quantum_bible, quantum_book}.

Shor's algorithm is as follows.
Two registers of $m$ and $n$ qubits, respectively, are initialized to the state $\ket{0}\ket{0}$.
The first register is put in a uniform superposition of all possible states:
\begin{equation}
\frac{1}{\sqrt{2^m}}\sum_{x=0}^{2^m-1}\ket{x}\ket{0}.
\label{shor_init}
\end{equation}
Next, $a^x \mod X$ is computed for all $x\in[0,2^m-1]$ and the result is stored in the second register, thus (\ref{shor_init}) becomes
\begin{equation}
\frac{1}{\sqrt{2^m}}\sum_{x=0}^{2^m-1}\ket{x}\ket{a^x \mod X}.
\label{shor_exp}
\end{equation}
This entangles the registers and encodes the period $r$ in the second register.
The second register is measured to sample a single point in the period; that is, a fixed $y\in[0,2^m-1]$ is selected, yielding
\begin{equation}
\frac{1}{z}\sum_{x \in A}\ket{x}\ket{y},
\label{shor_measure2}
\end{equation}
where $z$ is a normalization term that preserves unitarity and
\begin{equation*}
    A = \{x\in[0,2^m-1] : y=a^x \mod X\}.
\end{equation*}
However, although this captures the period in the first register, in general it is offset by a global phase $k$; that is, $A=\{k, k+r, k+2r, \dots \}$.
Performing the inverse quantum Fourier transform on the first register removes this offset and transforms the state to
\begin{equation}
\frac{1}{z}\left(e^{0}\ket{0}\ket{y}+e^{\frac{i2\pi rk}{m}}\ket{r}\ket{y}+e^{\frac{i4\pi rk}{m}}\ket{2r}\ket{y}+\dots\right).
\label{shor_qft}
\end{equation}
Finally, the first register is measured to extract a number $\hat{r}$ that is a multiple of $r$.
Applying the continued fractions algorithm (CFA) to $\hat{r}/2^m$ reduces the order of this fraction, yielding an estimate of $r$ we denote $\Tilde{r}$.
If $\Tilde{r}$ is even then by (\ref{factors}) we obtain our factor of $X$ with high probability \cite{shor, quantum_bible, quantum_book}.

The use of (random) measurement means Shor's algorithm is probabilistic, and consequently has a nontrivial chance of failure.
This happens if $\Tilde{r}$ is odd or if $\gcd(X,a^{\Tilde{r}/2}\pm1)$ is not a factor of $X$, in which case one repeats the experiment.
Shor showed that few iterations are needed to yield $r$ with high probability \cite{shor}.

The number of qubits in the first register, $m$, must be large enough to capture several periods of $a^x \mod X$.
Mermin suggests choosing $m \ge 2x$, where $x$ is the number of bits in $X$, to ensure success \cite{quantum_book}.
The number of qubits in the second register, $n$, must be large enough to encode the largest value of $a^x \mod X$, so $n=\lceil \log(X-1) \rceil$ suffices.

UNIQuE implements Shor's algorithm, defined as \texttt{shors(X, a, m, n)}.
$\texttt{X}\in\mathbb{Z}$ is the number to factor, $\texttt{a}\in\mathbb{Z}$ is co-prime to \texttt{X}, and $\texttt{m},\texttt{n}\in\mathbb{N}$ represent the number of qubits in the first and second quantum registers, respectively.
The details of this algorithm are shown in Algorithm~\ref{shors_code}.
Here $\ket{\phi}$ represents the first register, $\ket{\sigma}$ represents the second register, and $\ket{\tau}$ represents a third register which tracks entanglement.

\begin{figure}[t]
\begin{algorithm}[H]
    \caption{Shor's Algorithm}\label{shors_code}
    \begin{algorithmic}[1]
        \Require $\text{Integer } X, \text{ integer } a, \text{ integer } m, \text{ integer } n$
        \Procedure{Shors}{$X, a, m, n$}:
            \State $\ket{\phi} \gets \ket{1}\text{(ones vector) of length }2^m$
            \State $\ket{\sigma} \gets \ket{0}\text{of length }2^n$
            \State $\ket{\tau} \gets \ket{0}\text{of length }2^m$
            \For{$x \gets 0, 1, ..., 2^m-1$}
                \State $y \gets a^x\mod X$
                \State $\ket{\sigma}_y \gets \ket{\sigma}_y + 1$
                \State $\ket{\tau}_x = y$
            \EndFor
            \State $\ket{\sigma} \gets \text{normalize}(\ket{\sigma})$
            \State $\ket{\sigma} \gets \text{measure}(\ket{\sigma})$
            \State $i \gets \text{index of nonzero element of }\ket{\sigma}$
            \State $\ket{\phi}_{\ket{\tau} \neq i} \gets 0$
            \State $\ket{\phi} \gets \text{normalize}(\ket{\phi})$
            \State $\ket{\phi} \gets \text{IQFT}(\ket{\phi})$
            \State $\ket{\phi} \gets \text{measure}(\ket{\phi})$
            \State $r \gets \text{index of nonzero element of }\ket{\phi}$
            \State $r \gets \text{CFA}(r/2^m)$
            \If{$2\text{ divides }r$}
                \State \Return $\gcd(a^{r/2}-1, X),\text{ }\gcd(a^{r/2}+1, X)$
            \Else
                \State \Return $r$
            \EndIf
        \EndProcedure
    \end{algorithmic}
\end{algorithm}
\end{figure}

First, UNIQuE creates a ones array of size $2^m$ (line 2) and a zeros array of size $2^n$ (line 3).
Together, these represent (\ref{shor_init}).
UNIQuE also creates a third array of size $2^m$ (line 4) which does not not appear in Shor's algorithm but is needed to emulate entanglement between the two algorithm registers.
Next, $y = a^x \mod X$ is computed for all $x\in[0,2^m-1]$ (line 6); the second register records the frequency of each $y$ (line 7); and the third register emulates entanglement by mapping the indices $x$ to values $y$ (line 8).
Together, these represent (\ref{shor_exp}).
To compute (\ref{shor_measure2}), UNIQuE measures the second register to select one $y$ (line 10), and then uses the third register to eliminate all values of the first register which do not correspond to $y$ (line 12).
Next, the inverse quantum Fourier transform is computed on the first register (line 14) to get (\ref{shor_qft}), after which the first register is measured (line 15) and the CFA is applied to produce $\Tilde{r}$ (line 17).
If $\Tilde{r}$ is even then we compute and return $\gcd(X,a^{\Tilde{r}/2}\pm1)$ (line 19).

We tested this algorithm on four problems, given in Table~\ref{shors_table}.
These problems have special significance: $15$ was the first number factored on a quantum computer \cite{factor}; $35$ was factored by hundreds of people during IBM's 2021 quantum computing challenge \cite{IBM}; and both $8509=67\times127$ and $42781=179\times239$ are the product of two relatively large primes and thus represent prototypes for an RSA public-key.
Each number was factored ten times on UNIQuE; the average time per computation is recorded in the table along with the average accuracy across all computations.

\begin{table}[t]
    \centering
    \begin{tabular}{| c | c | c | c | c | c |}
        \hline
        $X$ & $a$ & $m$ & $n$ & Time (s) & Accuracy \\
        \hline
        15 & 7 & 8 & 4 & 0.0010 & 0.8 \\
        \hline
        35 & 13 & 8 & 5 & 0.0010 & 0.8 \\
        \hline
        8509 & 38 & 28 & 14 & 555.3564 & 0.9 \\
        \hline
        42781 & 10 & 28 & 16 & 1662.3580 & 0.8 \\
        \hline
    \end{tabular}
    \caption{Performance of Shor's algorithm.}
    \label{shors_table}
\end{table}

Notably, using 44 qubits and just over 27 minutes of computation time, UNIQuE successfully factored 42781 using Shor's algorithm.
This represents a significant achievement.
It is not possible to factor 42781 on any existing quantum computers today.
Although sufficiently large quantum computers exist (such as IBM's 1,121 qubit Condor processor \cite{condor}), current error rates ensure that a factoring algorithm will fail before it successfully terminates.
Quantum simulators have trouble factoring this number as well, for at least two potential reasons.
First, some simulators, like Intel-QS, require the user to build the quantum circuit for their algorithm themselves.
Determining the exact sequence of one and two qubit gates to factor 42781 is prohibitively difficult even for experienced users.
Second, once a gate sequence for the algorithm is obtained (manually or by a built in function), the size of the problem is too large to be simulated efficiently.

\begin{figure}[t]
    \centering
    \includegraphics[width=\figsize\linewidth]{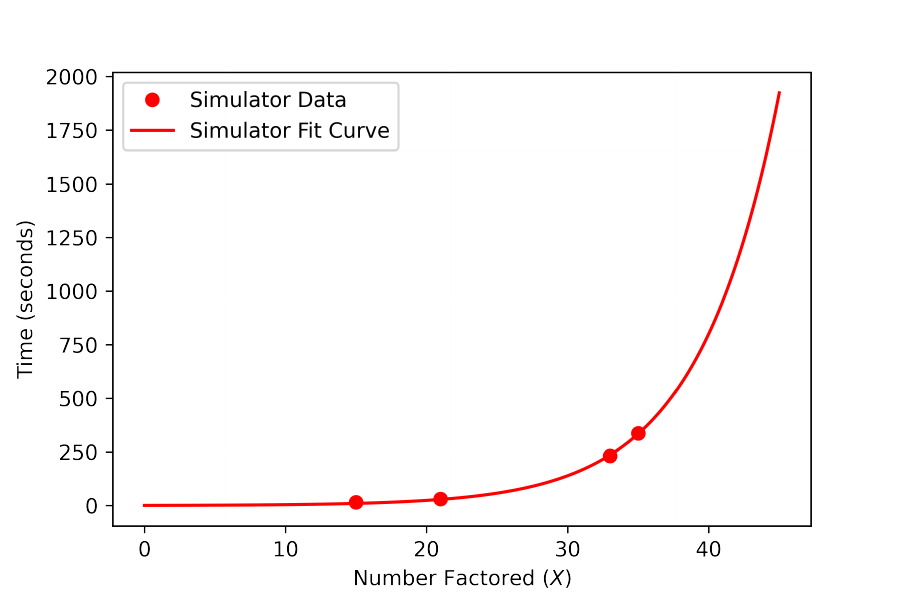}
    \caption{
        Time required for simulating Shor's algorithm on IBM Quantum as a function of $X$.
        By $X=45$ the time required is greater than 1662 seconds---the time it took UNIQuE to factor 42781.
    }
    \label{shors_interp_figure}
\end{figure}

To demonstrate this second point we tried to factor 42781 on IBM Quantum (formerly IBM Quantum Experience).
Our attempt ran on the IBM mainframe for over an hour (more than double UNIQuE's time), after which we terminated the program.
To estimate the time required for this problem we factored several smaller numbers (15, 21, 33, and 35) and fit a curve to the results (see Fig.~\ref{shors_interp_figure}).
The curve of best fit is
\begin{equation*}
    y \approx 0.729 \times 2^{0.253 n},
\end{equation*}
so we estimate that IBM Quantum will require $1.227 \times 10^{3258}$ seconds to factor the number 42781.
In other words, this system cannot solve this problem.

To summarize, using UNIQuE we emulated Shor's algorithm and factored a number too large for current quantum simulators and several quantum devices.
Current quantum devices that can attempt our problem will likely fail due to errors in the system.
Consequently, the emulator makes all the difference in studying the behavior of Shor's algorithm on difficult problems in the current noisy intermediate-scale quantum (NISQ) era.
Finally, on a user-friendliness note, UNIQuE allows users to execute Shor's algorithm without requiring them to build the quantum circuit themselves.

\section{Conclusion} \label{conclusion}

In this paper we introduce the Unconventional Noiseless Intermediate Quantum Emulator (UNIQuE), a software solution which performs quantum computations on a classical computer.
UNIQuE is a quantum computing \textit{emulator}, meaning that it abstracts entire quantum algorithms into optimized classical functions---in contrast to traditional quantum \textit{simulators}, which mimic each step of a quantum algorithm using matrix multiplication.

UNIQuE demonstrates computational advantage over the Intel Quantum Simulator (Intel-QS) in performing arithmetic operations, the quantum Fourier transform, and quantum phase estimation (QPE).
The most drastic advantage arises when generating increasingly precise phase estimations for a constant $U$ matrix; QPE runtime on Intel-QS scales exponentially with precision, while QPE runtime on UNIQuE is constant regardless of precision.
For all other operations tested both UNIQuE and Intel-QS scale exponentially with problem size, however UNIQuE scales with a lower coefficient, making a significant difference in tractability for many problems.
The computational advantage can be improved further through the use of sparse operations where possible (this is demonstrated with the arithmetic operations).

UNIQuE also demonstrates appreciable spatial advantage generally; and significant spatial advantage where sparse operations can be utilized.
Here striking examples include a 40 qubit QPE problem that would require a simulator to store approximately $10^{6300000}$ values (and more are required to compute the solution), a 44 qubit Shor's algorithm with similar complexity, and a 20 by 20 qubit exponentiation problem with astronomical simulation requirements.

Additionally, we used UNIQuE to evaluate Shor's algorithm on a nontrivial problem.
This demonstrates that UNIQuE can be leveraged for practical quantum algorithms.
In other words, UNIQuE allows researchers to study the ideal behavior of quantum algorithms on problems intractable in the current NISQ era.

Finally, UNIQuE is not a finished product---instead it is an open-source piece of software ready to be modified and extended.
There are several additions that could be made in the future.
For example, other notable algorithms could be emulated directly, including the Deutsch-Jozsa algorithm \cite{deutsch} and Grover's quantum search algorithm \cite{grover}, as well as emerging algorithms such as Yamakawa and Zhandry's recent algorithm for learning-with-errors on unstructured problems \cite{learn_error}.
Additionally, we hope that the functions offered by UNIQuE will be useful building blocks for future algorithm experimentation and development.

\appendix

\section{UNIQuE Documentation} \label{documentation}

UNIQuE implements eight basic operations: \texttt{normalize}, \texttt{add}, \texttt{multiply}, \texttt{exponentiate}, \texttt{qft}, \texttt{inv\_qft}, \texttt{qpe}, and \texttt{measure}.
It also implements a variation on the first four operations that employs sparse matrices to achieve greater spatial savings, which operations are \texttt{normalize\_sparse}, \texttt{add\_sparse}, \texttt{multiply\_sparse}, \texttt{exponentiate\_sparse}.
These operations are discussed below after their non-sparse counterparts.
Finally, UNIQuE implements Shor's algorithm, \texttt{shors}, the celebrated quantum factoring algorithm.
Note that because we are modeling state vectors of qubits, in all operations below it is assumed (but not required) that $N=2^n$ for some positive integer $n$. 

\begin{enumerate}
    \item \texttt{normalize(x)}:
    takes as input a numpy array \texttt{x} of length $N$, computes the normalization of \texttt{x} according to (\ref{normalize}), and returns a numpy array of length $N$ containing the result.
    
    \item \texttt{add(a, b)}:
    accepts two numpy arrays \texttt{a} and \texttt{b} of sizes $N_1$ and $N_2$, respectively, computes the sum $\texttt{a}+\texttt{b}$ using Algorithm~\ref{add}, and returns a numpy array \texttt{c} of size $2\times\max(N_1 + N_2)$ containing the result.
    
    \item \texttt{multiply(a, b)}:
    accepts two numpy arrays \texttt{a} and \texttt{b} of sizes $N_1$ and $N_2$, respectively, computes the product $\texttt{a} \times \texttt{b}$ using Algorithm~\ref{multiply}, and returns a numpy array \texttt{c} of length $N_1\times N_2$ containing the result.
    
    \item \texttt{exponentiate(a, b)}:
    accepts two numpy arrays \texttt{a} and \texttt{b} of sizes $N_1$ and $N_2$, respectively, computes $\texttt{a}^{\texttt{b}}$ using Algorithm~\ref{exponentiate}, and returns a numpy array \texttt{c} of length $N_1^{N_2}$ containing the result.
    Note that for this operation the ordering of \texttt{a} and \texttt{b} matters, as $\texttt{a}^{\texttt{b}}\neq\texttt{b}^{\texttt{a}}$ in general.
    
    \item \texttt{normalize\_sparse(x)}:
    performs the same operation as the \texttt{normalize} function, that is, it performs (\ref{normalize}) on its input.
    However, rather than using numpy arrays, this function uses the \texttt{scipy.sparse.dok\_matrix} framework.
    It accepts a dok\_matrix as input and returns a normalized dok\_matrix for the output.
    
    \item \texttt{add\_sparse(a, b)}:
    operates identically to the \texttt{add} function above, however \texttt{a}, \texttt{b}, and \texttt{c} are dok\_matrix objects rather than numpy arrays.
    
    \item \texttt{multiply\_sparse(a, b)}:
    operates identically to the \texttt{multiply} function above, however \texttt{a}, \texttt{b}, and \texttt{c} are dok\_matrix objects rather than numpy arrays.
    
    \item \texttt{exponentiate\_sparse(a, b)}:
    operates identically to the \texttt{exponentiate} function above, however \texttt{a}, \texttt{b}, and \texttt{c} are dok\_matrix objects rather than numpy arrays.
    
    \item \texttt{qft(x)}:
    accepts a numpy array \texttt{x} of length $N$, uses \texttt{scipy.fftpack.ifft}---the inverse discrete Fourier transform implemented by scipy---to classically compute the quantum Fourier transform, applies the \texttt{normalize} function to preserve the vector's unitary property, and returns a numpy array of length $N$ containing the result.
    See Algorithm~\ref{qft_code}.
    
    \item \texttt{inv\_qft(x)}:
    accepts a numpy array \texttt{x} of length $N$, uses \texttt{scipy.fftpack.fft}---the discrete Fourier transform implemented by scipy---to classically compute the inverse quantum Fourier transform, applies the \texttt{normalize} function to preserve the vector's unitary property, and returns a numpy array of length $N$ containing the result.
    
    \item \texttt{qpe(U, phi, n)}:
    takes three inputs: an $M\times M$ unitary matrix \texttt{U}, an $M\times1$ eigenvector \texttt{phi} in the form of a numpy array, and an integer \texttt{n} which specifies the number of qubits of precision to use for the output of the function.
    It is required that $M=2^m$ for some positive integer $m$.
    The function finds all of the eigenvalues and eigenvectors of \texttt{U} using \texttt{numpy.linalg.eig(U)}, and determines which eigenvalue corresponds to the eigenvector \texttt{phi}.
    Because the eigenvalue is of the form $e^{2\pi i\theta}$, $\theta$ is extracted, and the integer $r\in[0,N]$ is found such that $r/N$ is the closest possible approximation of $\theta$, where $N=2^n$.
    This value $r$ is encoded into a state vector (a dok\_matrix specifically) of size $N$ and returned as the output of the function.
    See Algorithm~\ref{qpe_code}.
    
    \item \texttt{measure(x, return\_index=False)}:
    takes a numpy array \texttt{x} of length $N$ as input and returns a numpy array of the same size with a single nonzero entry, which entry is $1$.
    The probability that any given index in the output will hold the value $1$ is given by the value of the corresponding entry of \texttt{x} squared, as discussed in Section~\ref{math}.
    This random selection is made using the \texttt{numpy.random.choice} function.
    If \texttt{return\_index=True} then the index of the nonzero state is also returned.

    \item \texttt{shors(X, a, m, n)}:
    the inputs to this function are as follows: $\texttt{X}\in\mathbb{Z}$ is the number to factor, $\texttt{a}\in\mathbb{Z}$ is co-prime to \texttt{X}, and $\texttt{m},\texttt{n}\in\mathbb{N}$ represent the number of qubits in the first and second quantum registers, respectively.
    It returns an estimate of the factors of \texttt{X}.
    See Section~\ref{application} for a more detailed discussion of this operation.
\end{enumerate}

\section{Additional Arithmetic Algorithms} \label{arithmetic_algs_appendix}

Below are the general algorithms for multiplication and exponentiation.
See also Section~\ref{arithmetic_math}, and in particular Algorithm~\ref{add}.

\newpage

\begin{figure}[htbp]
\begin{algorithm}[H]
    \caption{Multiplication}\label{multiply}
    \begin{algorithmic}[1]
        \Require $\text{State vector } \ket{\alpha},\text{ state vector } \ket{\beta}$
        \Procedure{Multiply}{$\ket{\alpha}, \ket{\beta}$}:
            \State $N_{\alpha} \gets \text{length}(\ket{\alpha})$
            \State $N_{\beta} \gets \text{length}(\ket{\beta})$
            \State $\ket{\gamma} \gets \ket{0}\text{of length } N_{\alpha} \times N_{\beta}$
            \For{$i \gets 0, 1, ..., N_{\alpha}-1$}
                \For{$j \gets 0, 1, ..., N_{\beta}-1$}
                    \State $\ket{\gamma}_{i \times j} \pluseq \ket{\alpha}_i \times \ket{\beta}_j$
                \EndFor
            \EndFor
            \State $\ket{\gamma} \gets \text{normalize}(\ket{\gamma})$
            \State \Return{$\ket{\gamma}$}
        \EndProcedure
    \end{algorithmic}
\end{algorithm}
\end{figure}

\begin{figure}[htbp]
\begin{algorithm}[H]
    \caption{Exponentiation}\label{exponentiate}
    \begin{algorithmic}[1]
        \Require $\text{State vector } \ket{\alpha},\text{ state vector } \ket{\beta}$
        \Procedure{Exponentitate}{$\ket{\alpha}, \ket{\beta}$}:
            \State $N_{\alpha} \gets \text{length}(\ket{\alpha})$
            \State $N_{\beta} \gets \text{length}(\ket{\beta})$
            \State $\ket{\gamma} \gets \ket{0}\text{of length } N_{\alpha}^{N_{\beta}}$
            \For{$i \gets 0, 1, ..., N_{\alpha}-1$}
                \For{$j \gets 0, 1, ..., N_{\beta}-1$}
                    \State $\ket{\gamma}_{i^j} \pluseq \ket{\alpha}_i \times \ket{\beta}_j$
                \EndFor
            \EndFor
            \State $\ket{\gamma} \gets \text{normalize}(\ket{\gamma})$
            \State \Return{$\ket{\gamma}$}
        \EndProcedure
    \end{algorithmic}
\end{algorithm}
\end{figure}

\bibliography{references.bib}

\end{document}